\documentclass[aps,prd,floats,floatfix,twoside,preprintnumbers,superscriptaddress,nofootinbib]{revtex4}

\usepackage{graphicx} 
\usepackage{dcolumn}
\usepackage{amssymb}
\usepackage{amsmath}
\usepackage[hidelinks]{hyperref}
\usepackage{color}
\usepackage[utf8]{inputenc}

\newcommand{\qbar}{\bar{q}}

\newcommand{\hats}{\hat{s}} 
\newcommand{\hatt}{\hat{t}} 
\newcommand{\hatu}{\hat{u}} 
\newcommand{\hatz}{\hat{z}}

\begin{document}

\preprint{YITP-18-66}

\date{\today}

\title{Single spin asymmetries in ultra-peripheral $p^\uparrow A$ collisions}
\author{Sanjin Beni\' c\footnote{On leave of absence from: Department of Physics, Faculty of Science, University of Zagreb, Bijenička c. 32, 10000 Zagreb, Croatia}}
\author{Yoshitaka Hatta}
\affiliation{Yukawa Institute for Theoretical Physics,
Kyoto University, Kyoto 606-8502, Japan}
\begin{abstract} 
We suggest inclusive hadron production in  ultra-peripheral proton-nucleus collisions (UPCs) $p^\uparrow A \to h AX$ as a new channel to investigate single spin asymmetries (SSAs), in particular, to test the assumed dominance of the contribution from twist-three fragmentation  functions. 
The UPC cross sections are obtained by considering the photoproduction limit of semi-inclusive deep inelastic scattering (SIDIS). In particular, we find simple formulas for the polarized UPC cross sections in the collinear twist-three framework.
 We then numerically calculate the fragmentation contribution to SSA in $p^\uparrow A \to \pi A X$ at $\sqrt{s} = 200$ GeV and find  a few percent asymmetry in the forward region.
\end{abstract}


\maketitle 

\section{Introduction}

Single spin asymmetries (SSAs) are an important tool to understand the spin and tomographic structures of the  polarized proton \cite{DAlesio:2007bjf,Boer:2011fh,Aschenauer:2015eha,Boglione:2015zyc}. For example, inclusive hadron production in $p^\uparrow p$ collisions, where $p^\uparrow$ denotes the transversely polarized proton, is one of the simplest ways to study SSA. Experimentally, largest asymmetries in $p^\uparrow p$ were found in the forward region \cite{Adams:2003fx,Abelev:2008af,Adamczyk:2012xd,Lee:2007zzh}. At high $P_{h\perp}$ of the produced hadron, the collinear framework is appropriate and the asymmetry essentially becomes a probe of the twist-3 effects, see \cite{Pitonyak:2016hqh} for a recent review. In particular, recent fits seem to point to the twist-3 fragmentation functions as the dominant contribution to SSA \cite{Kanazawa:2014dca,Gamberg:2017gle}.

In this work we suggest to study SSA for inclusive hadron production in ultra-peripheral $pA$ collisions (UPCs).
Currently, RHIC conducts experiments with polarized protons on a nuclear target \cite{Dilks:2016ufy,Aidala:2017cnz}. Moreover, polarized projectile protons may be possible at the LHC in the fixed target mode (AFTER@LHC) \cite{Brodsky:2012vg,Lansberg:2016gwm}. It has been shown that the dependence of SSA on the atomic number $A$ (or the nuclear charge $Z$) is sensitive to the gluon saturation effect of the unpolarized nucleus \cite{Boer:2006rj,Kang:2011ni,Hatta:2016wjz,Hatta:2016khv}, and this fact can be used to discriminate different mechanisms of SSAs. However, as $Z$ becomes large, the cross section of UPCs is parametrically similar to that of purely hadronic cross sections, so the contribution to the asymmetry can be large and should be taken into account. In principle, and actually in practice at both RHIC and the LHC, experimentalists can single out ultra-peripheral events since the nuclei stays intact after the collision. 
Since UPC is potentially theoretically cleaner than a purely hadronic collision, by separating it out in an experiment we may be able to get new constraints on the various nonperturbative functions responsible for SSAs.

It should be mentioned that SSA in UPCs has been previously studied in a few specific channels. 
Ref.~\cite{Mitsuka:2017czj} suggested that the strong $Z$-dependence of SSA for very forward, very low-$P_T$ neutrons observed by the PHENIX collaboration \cite{Aidala:2017cnz} could be attributed to UPCs. In \cite{Goncalves:2017fkt}, SSA for quarkonia production in $pA$ UPC was proposed as a probe of the gluon Sivers function.  
In this work we instead use the collinear twist-3 framework and calculate the polarized and the unpolarized cross section for $p^{\uparrow}(p,S)  A \to h(P_h)  X$ production, with the nuclei described by an equivalent photon flux. 
In this setup, the process essentially becomes a $p\gamma$ collision, similar to deep inelastic scattering (DIS) in the photo-production limit. Our calculation will therefore rely on the existing results for SSA in semi-inclusive DIS (SIDIS) 
$e (l) p^\uparrow (P,S) \to e(l') h(P_h) X$ coming from the twist-3 quark-gluon-quark distribution \cite{Eguchi:2006qz,Eguchi:2006mc}, from twist-3 fragmentation function \cite{Eguchi:2006qz,Yuan:2009dw,Kang:2010zzb,Metz:2012ct,Kanazawa:2013uia} and from twist-3 gluon distribution \cite{Beppu:2010qn}. Keeping the detected hadron   transverse momentum $P_{h\perp}$ to be large, we can take the photoproduction limit $Q^2 \to 0$, where $q^2 = - Q^2$ is the photon virtuality. In this limit the SIDIS results collapse to a remarkably simple form.

We consider inclusive pion production, $p^\uparrow A \to \pi X$, at $\sqrt{s} = 200$ GeV and numerically calculate the asymmetry assuming that the forward region is dominated by the twist-3 fragmentation contribution. We use the recent extraction of the transversity and the pion twist-3 fragmentation functions from \cite{Kang:2015msa}. We only consider the so-called direct photon process. The contribution from the resolved photon process may not be negligible in the forward region of the polarized proton (backward region of the photon)  \cite{Helenius:2017aqz}, but this may be eliminated by requiring additional cuts in the final state.

This paper is organized as follows: in Sec.~\ref{sec:ind} we give the main formula for the spin independent cross section as well as a brief review of the relevant kinematics. The formulas for the spin dependent cross section are given in Sec.~\ref{sec:dep}. In Sec.~\ref{sec:num} we present the numerical results and in Sec.~\ref{sec:conc} we give our conclusions.

\section{Spin independent cross section}
\label{sec:ind}

In this Section we will take the known SIDIS $e (l) p (P) \to e(l') h(P_h) X$ cross section as calculated in \cite{Meng:1991da}, see also \cite{Eguchi:2006mc}, and adapt it to the $Q^2 \to 0$ limit suitable for describing UPCs. 
The SIDIS kinematic variables are defined as
\begin{equation}
S_{ep} = (P + l)^2 \,, \quad x_{ bj} = \frac{Q^2}{2P \cdot q} \,, \quad Q^2 = -q^2 = (l - l')^2 \,, \quad z_f = \frac{P\cdot P_h}{P\cdot q}\,,
\end{equation}
where $S_{ep}$ is the lepton-proton center of mass energy, $x_{bj}$ is the Bjorken-$x$, $Q^2$ is the photon virtuality and $z_f$ is a scaling variable.
SIDIS is calculated in the hadron frame, but since the formula are valid in any frame where $\boldsymbol{P}$ and $\boldsymbol{q}$ are collinear it can be used also in the photoproduction limit $Q^2 \to 0$.

In order to take the $Q^2 \to 0$ limit we need several steps.
First we recover the final lepton and hadron phase space from the conventional SIDIS integration variables
\begin{equation}
\frac{d^3 P_h}{(2 P_h^0)(2\pi)^3}\frac{d^3 l'}{(2 l'^0)(2\pi)^3} = \frac{4 z_f Q^2}{{(4\pi)^6 x_{bj}^2 S_{ep}}}d x_{bj} dQ^2 dz_f d q_T^2 d\phi d\chi\,.
\end{equation}
Here $q_T = P_{h\perp}/z_f$, $\phi$ and $\chi$ are lepton and hadron azimuthal angles, respectively. Then, we can remove the final lepton phase space. Next, we simplify the coefficients that enter into the cross section (see e.~g. Eq.~(52) in \cite{Eguchi:2006mc})
\begin{equation}
\mathcal{A}_k = \frac{L_{\mu\nu}\mathcal{V}_k^{\mu\nu}}{Q^2}\,,
\end{equation}
where $L_{\mu\nu}$ is the leptonic tensor and $\mathcal{V}_k^{\mu\nu}$ are defined in Eq.~(49) of Ref~\cite{Eguchi:2006mc}.
For a real photon we replace $L_{\mu\nu}$ with
\begin{equation}
\frac{e^2}{q^4} L_{\mu\nu} \to \sum_\lambda\epsilon_\mu(q,\lambda) \epsilon^*_{\nu}(q,\lambda) \to - g_{\perp \mu\nu}\,,
\end{equation}
where $g_{\perp \mu\nu}$ is the transverse part of $g_{\mu\nu}$.
This simplifies $\mathcal{A}_k$ so that $\mathcal{A}_1 = 2Q^2/e^2$, $\mathcal{A}_2 = -2 Q^2/e^2$ while the rest $\mathcal{A}_k = 0$ with $k = 3,4$.
Finally, we also replace the incident flux pre-factor $1/(2S_{ep})$ with $1/(2 S)$ where  $S \equiv (q + P)^2$.
For later convenience, we introduce the remaining Mandelstam variables: $T \equiv (P - P_h)^2$ and $U \equiv (q - P_h)^2$. Likewise, on the parton level we will use: $\hats \equiv (q + xP)^2 = xS$, $\hatt \equiv (xP - P_h/z)^2 = xT/z$, $\hatu \equiv (q - P_h/z)^2 = U/z$, where $x$ is the light-cone momentum fractions for the initial state parton and $z$ is the momentum fraction of the outgoing hadron.

Using the above replacements in the $Q^2\to 0$ limit we obtain the unpolarized $p\gamma\to h X$ cross section from Eq.~(54) in \cite{Eguchi:2006mc}. We find
\begin{equation}
\begin{split}
\frac{d\sigma}{d^2 P_{h\perp} dy_h} &= \frac{8\alpha_{em}\alpha_s}{z_f S}\int_{x_{\rm min}}^1 \frac{dx}{x}\int_{z_{\rm min}}^1 \frac{dz}{z}\delta\left(q_T^2 + \left(1-\frac{1}{\hatz}\right)\hats\right)\\
&\times\sum_{a} e_a^2\left[f_a(x,\mu^2) D_{h/a}(z,\mu^2)\hat{\sigma}^{qq}_1 + f_g(x,\mu^2) D_{h/a}(z,\mu^2)\hat{\sigma}^{qg}_1 + f_a(x,\mu^2) D_{h/g}(z,\mu^2)\hat{\sigma}^{gq}_1 \right]\\
&= \frac{8\alpha_{em}\alpha_s}{S}\int_{x_{\rm min}}^1 \frac{dx}{x}\int_{z_{\rm min}}^1 \frac{dz}{z^2}\delta\left(\hats + \hatt + \hatu\right)\\
&\times\sum_{a} e_a^2\left[f_a(x,\mu^2) D_{h/a}(z,\mu^2)\hat{\sigma}^{qq}_1 + f_g(x,\mu^2) D_{h/a}(z,\mu^2)\hat{\sigma}^{qg}_1 + f_a(x,\mu^2) D_{h/g}(z,\mu^2)\hat{\sigma}^{gq}_1 \right]\,,
\label{eq:xqqgunpol}
\end{split}
\end{equation}
where $f_a(x,\mu^2)$ are the parton distribution functions of a particular flavor $a$, while $D_{h/a}(z,\mu^2)$ are the parton-to-hadron unpolarized fragmentation functions evaluated at the scale $\mu^2$. In the numerical calculations performed in Sec.~\ref{sec:num} we use $\mu^2 = P_{h\perp}^2$. 
We have used $d^3 P_h/(2 P_h^0) = d^2 P_{h\perp} dy_h/2$ to rewrite the outgoing hadron phase space in terms of its transverse momenta $P_{h\perp}$ and the rapidity $y_h$.
Also, here and in the rest of the paper we have $x_{\rm min} = \frac{z_f}{1-z_f}\frac{q_T^2}{S} = -U/(T + S)$ and $z_{\rm min} = z_f\left(1 + \frac{q_T^2}{S}\right) = -(T + U)/S$ and $\hatz = z_f/z$. To get the second line of \eqref{eq:xqqgunpol} we have used that $\hatz = -\hatt/\hats$ and $q_T^2 = \hats \hatu /\hatt$.
The relevant hard factors, see Eqs.~(57)-(59) in \cite{Eguchi:2006mc}, in the $Q^2 \to 0$ limit are
\begin{equation}
\begin{split}
&\hat{\sigma}_1^{qq} = 2 C_F \left(\frac{1}{\hatz} \frac{\hats}{q_T^2} + \hatz \frac{q_T^2}{\hats}\right) = 2C_F \frac{1 + (1-\hatz)^2}{1-\hatz} = -2C_F \left(\frac{\hats}{\hatu} + \frac{\hatu}{\hats}\right)\,,\\
&\hat{\sigma}_1^{qg} = \frac{1}{\hatz^2}\frac{\hats}{q_T^2}-2 = \frac{1}{1-\hatz}\frac{(1-\hatz)^2 + \hatz^2}{\hatz} = \frac{\hatt}{\hatu} + \frac{\hatu}{\hatt}\,,\\
&\hat{\sigma}_1^{gq} = 2 C_F \left(\frac{1-\hatz}{\hatz^2} \frac{\hats}{q_T^2} + \frac{\hatz^2}{1-\hatz} \frac{q_T^2}{\hats}\right) = 2C_F \frac{1 + \hatz^2}{\hatz} = -2C_F \left(\frac{\hats}{\hatt} + \frac{\hatt}{\hats}\right)\,,
\end{split}
\label{eq:qqgunpol}
\end{equation}
where $C_F = (N_c^2  -1)/(2 N_c)$. In the second equality in Eq.~\eqref{eq:qqgunpol} we used the $\delta$-function constraint in \eqref{eq:xqqgunpol}.

We take $q^\mu = (\omega,0,0,-\omega)$, and $P^\mu = (\sqrt{s}/2,0,0,\sqrt{s}/2)$, where $\omega$ is the photon energy and $s$ is the center of mass energy of the $pA$ collision per nucleon. We get $S = 2\omega \sqrt{s}$.
To get the cross section in $pA$ UPC we need to multiply the above $p\gamma\to h X$ cross sections by the photon flux $dN/d\omega$
\begin{equation}
\frac{d\sigma_{pA}}{d^2 P_{h\perp} dy_h} = \int_0^\infty d\omega \frac{dN}{d\omega} \frac{d\sigma}{d^2 P_{h\perp} dy_h}\,,
\label{eq:unpA}
\end{equation}
where
\begin{equation}
\frac{dN}{d\omega} = \frac{2Z^2\alpha_{em}}{\pi \omega}\left[\xi K_0(\xi)K_1(\xi)-\frac{\xi^2}{2}(K_1^2(\xi) - K_0^2(\xi))\right]\,,
\label{eq:pfl}
\end{equation}
with $\xi = \omega \frac{R_p + R_A}{\gamma}$ and $\gamma = \frac{\sqrt{s}}{2 M_N}$, where $R_p$ and $R_A$ are the charge radii of the proton and the nuclei and $M_N$ is the proton mass.

We close this section with a comment on the kinematics of the UPC channel. We have $z_f = (p\cdot P_h)/(p\cdot q) = \frac{1}{2}\frac{P_{h\perp}}{\omega}e^{-y_h}$, where $y_h$ is the rapidity of the hadron.
The integration over $x$ starts from $x_{\rm min}$. Requiring that $x_{\rm min} > 0$ leads to a condition $1-z_f >0$ which gives $\omega > P_{h\perp} e^{-y_h}/2$. Requiring $x_{\rm min}<1$ leads to a stronger condition
\begin{equation}
\omega > \omega_{\rm min} \equiv \frac{\frac{1}{2}P_{h\perp}e^{-y_h}}{1-\frac{P_{h\perp}}{\sqrt{s}}e^{y_h}}\,.
\end{equation}
Finally, note that going to arbitrarily forward region is not possible because the denominator has to be positive. Going to arbitrarily backward region brings a strong suppression on the cross section because the photon flux drops exponentially with large $\omega$.

\section{Spin dependent cross section}
\label{sec:dep}

In this Section we write down the main analytical formulas for the spin dependent $p^\uparrow \gamma \to \pi X$ cross section. The formula for the $p^\uparrow A \to \pi X$ cross section then follows by multiplying with the photon flux \eqref{eq:pfl}, as was done for the unpolarized cross section in Eq.~\eqref{eq:unpA}.
The formulas we obtained cover the complete twist-3 collinear contributions that includes the twist-3 quark-gluon correlations, the twist-3 gluon correlations as well as the twist-3 fragmentation contribution. As in the previous section, we use the already known SIDIS results and take the $Q^2 \to  0$ limit. The twist-3 quark-gluon and the twist-3 gluon contributions are collected in the Appendix \ref{app:qg3g}, while the twist-3 fragmentation contribution is elaborated in the following.

The twist-3 fragmentation contribution to the polarized SIDIS cross section was calculated in \cite{Eguchi:2006qz,Yuan:2009dw,Kang:2010zzb,Metz:2012ct,Kanazawa:2013uia} - we will use the notation in \cite{Kanazawa:2013uia}. 
The relevant cross section, see Eq.~(69) in \cite{Kanazawa:2013uia}, in the $Q^2 \to 0$ limit is
\begin{equation}
\begin{split}
& \frac{d\Delta\sigma_{\rm frag}}{d^2 P_{h\perp} d y_h} = \frac{4 \alpha_{em}\alpha_s M_h}{S}\sin(\Phi_S - \chi)\sum_{k=1,2}\tilde{\mathcal{A}}_k\int_{x_{\rm min}}^1 \frac{dx}{x}\int_{z_{\rm min}}^1 \frac{dz}{z^2}\delta\left(\hats + \hatt + \hatu\right)\sum_{a} e_a^2 h_1^a(x,\mu^2)\Bigg[-\frac{H^a(z,\mu^2)}{z}\Delta \hat{\sigma}_k^1\\
& + 2\frac{d}{d(1/z)}\left(\frac{H_1^{\perp (1),a}(z,\mu^2)}{z}\right)\Delta\hat{\sigma}^2_k + 2H_1^{\perp (1),a}(z,\mu^2)\Delta\hat{\sigma}_k^3 - 4\int_z^\infty \frac{dz'}{z'^2}P\left(\frac{1}{1/z - 1/z'}\right)\mathrm{Im} \hat{H}^a_{FU}(z,z',\mu^2)\Delta \hat{\sigma}^4_k\Bigg] \,,
\end{split}
\label{eq:xsf}
\end{equation}
where $\Phi_S$ is the azimuthal angle of the proton spin, $M_h$ is hadron mas, $\tilde{\mathcal{A}}_1 = 2$ and $\tilde{\mathcal{A}}_2 = -2$. Here $h_1^a(x,\mu^2)$ is the quark transversity, and $H^a(z,\mu^2)$, $H_1^{\perp (1),a}(z,\mu^2)$ and  $\mathrm{Im} \hat{H}^a_{FU}(z,z',\mu^2)$ are the so-called intrinsic, kinematical (first $k_\perp$-moment of the Collins function \cite{Collins:1992kk}), and the dynamical twist-3 fragmentation functions \cite{Ji:1993vw}, respectively, evaluated at the scale $\mu^2$, for which we adopted a notation in e.~g.~\cite{Kanazawa:2015ajw,Gamberg:2017gle}. In the numerical calculations in Sec.~\ref{sec:num} we use $\mu^2 = P_{h\perp}^2$. The correspondence to the notation in \cite{Kanazawa:2013uia} is established as (omitting the quark flavor index)
\begin{equation}
H(z,\mu^2) = - \frac{M_N}{M_h} \hat{e}_{\bar{1}}(z,\mu^2) \,, \qquad 
H_1^{\perp (1)}(z,\mu^2) = \frac{M_N}{2 M_h} \mathrm{Im}\tilde{e}(z,\mu^2) \,, \qquad \mathrm{Im}\hat{H}_{FU}(z,z',\mu^2) = \frac{M_N}{2 M_h}\mathrm{Im} \hat{E}_F(z',z,\mu^2)\,.
\label{eq:Hs}
\end{equation}
The relevant hard factors, see Eqs.~(72)-(102) in \cite{Kanazawa:2013uia}, in the $Q^2 \to 0$ limit are
\begin{equation}
\begin{split}
&\frac{\Delta \hat{\sigma}^1_1}{q_T} = - \frac{4C_F}{q_T^2}(-2 + 3 \hatz) = 4C_F\frac{\hatt}{\hats\hatu}\left(2+3\frac{\hatt}{\hats}\right)\,,\\
&\frac{\Delta \hat{\sigma}^1_2}{q_T} = -\frac{\Delta \hat{\sigma}^1_2}{q_T} = \frac{\Delta \hat{\sigma}^4_2}{2 q_T} = - \frac{8 C_F \hatz}{q_T^2} = 8C_F \frac{\hatt^2}{\hats^2 \hatu}\,,\\
&\frac{\Delta \hat{\sigma}^2_1}{q_T} = - \frac{4 C_F}{q_T^2} = -4C_F \frac{\hatt}{\hats \hatu}\,,\\
&\Delta \hat{\sigma}^2_2 = 0\,,\\
&\frac{\Delta \hat{\sigma}^3_1}{q_T} = \frac{4 C_F }{q_T^2}(-1+3\hatz) = -4C_F\frac{\hatt}{\hats\hatu}\left(1 + 3\frac{\hatt}{\hats}\right)\,,\\
&\frac{\Delta \hat{\sigma}^4_1}{q_T} = \frac{2 C_F}{q_T^2}(-1+3\hatz) + \frac{2}{q_T^2}\frac{1}{\hatz - \hatz'}\left(-\frac{1}{N_c} + \hatz N_c\right) = -2C_F\frac{\hatt}{\hats\hatu}\left(1 + 3\frac{\hatt}{\hats}\right) + \frac{2}{z(1/z - 1/z')}\frac{1}{\hatu}\left(\frac{1}{N_c} + N_c \frac{\hatt}{\hats}\right)\,.\\
\end{split}
\end{equation}

The twist-3 fragmentation functions are not independent, rather they satisfy a set of relations
\begin{equation}
\begin{split}
&H^a(z,\mu^2) = -2z H^{\perp (1),a}_1 (z,\mu^2) - 2z\int_z^\infty \frac{dz'}{z'^2}P\left(\frac{1}{1/z' - 1/z}\right)\mathrm{Im} \hat{H}^a_{FU}(z,z',\mu^2)\,,\\
& -H^a(z,\mu^2) = \frac{d}{d(1/z)}\left(\frac{H^{\perp (1),a}_1 (z,\mu^2)}{z}\right) + \frac{2}{z}\int_z^\infty \frac{dz'}{z'^2}P\left(\frac{1}{(1/z - 1/z')^2}\right)\mathrm{Im} \hat{H}^a_{FU}(z,z',\mu^2)\,.
\end{split}
\label{eq:ft3}
\end{equation}
where the first equation is called the QCD equation of motion relation \cite{Kang:2010zzb,Metz:2012ct}, while the second is the Lorentz invariance relation \cite{Kanazawa:2015ajw}.
Using the relations \eqref{eq:ft3} we find two simplifications in the cross section. First, they allow us to completely remove the terms in the cross section \eqref{eq:xsf} containing integrals over $z'$. As a side comment, note that such a simplification is specific to the $Q^2\to 0$ limit of the SIDIS cross section: for a general $Q^2 \neq 0$, and owing to a more complicated $z'$ dependence of the SIDIS hard factors, \eqref{eq:ft3} are insufficient to eliminate the $z'$ integrals.  Second, we find that the contribution with $k = 2$ vanishes (similar as in the case of the twist-3 quark-gluon and the twist-3 gluon contribution), so that the final expression for the cross section becomes
\begin{equation}
\begin{split}
& \frac{d\Delta\sigma_{\rm frag}}{d^2 P_{h\perp} d y_h} =  \frac{8 M_h P_{h\perp}\alpha_{em}\alpha_s}{S}\sin(\Phi_S - \chi)\int_{x_{\rm min}}^1 \frac{dx}{x}\int_{z_{\rm min}}^1 \frac{dz}{z^3}\delta\left(\hats + \hatt + \hatu\right)\\
&\times\sum_{a} e_a^2 h_1^a(x,\mu^2)\left[\left(H_1^{\perp (1),a}(z,\mu^2) - z \frac{d H_1^{\perp (1),a}(z,\mu^2)}{dz}\right)\Delta \hat{\sigma}_1 + \frac{H^a(z,\mu^2)}{z} \Delta \hat{\sigma}_2\right] \,,
\end{split}
\label{eq:xsf2}
\end{equation}
where
\begin{equation}
\begin{split}
&\Delta \hat{\sigma}_1 \equiv \frac{4}{N_c}\frac{1}{\hatt}\,,\\
&\Delta\hat{\sigma}_2 \equiv -\frac{2}{\hatu}\left(N_c + \frac{1}{N_c}\frac{\hats - \hatu}{\hatt}\right)\,.
\end{split}
\end{equation}

\section{Numerical results}
\label{sec:num}

\begin{figure}
  \begin{center}
  \includegraphics[scale = 0.18]{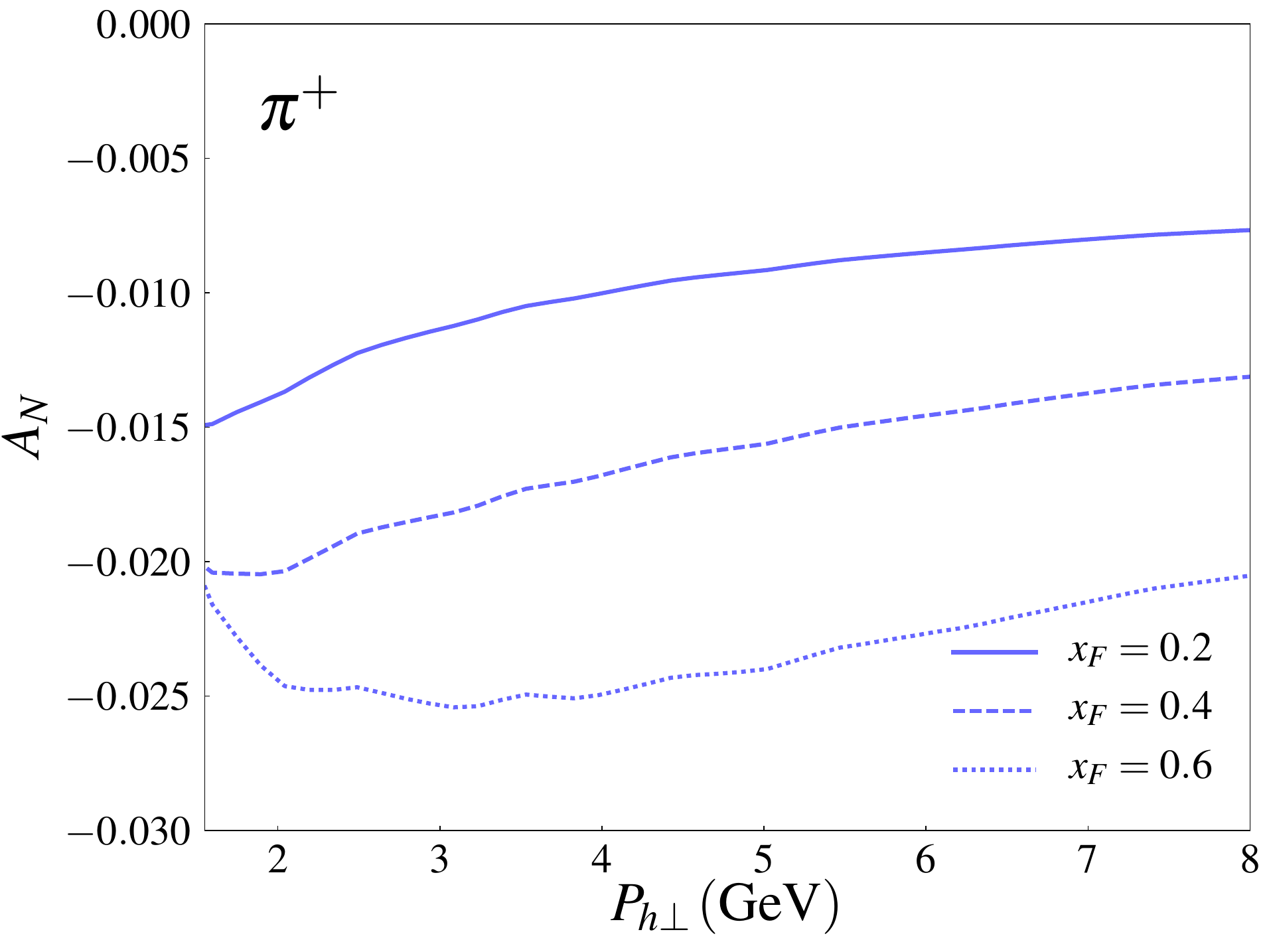}
    \includegraphics[scale = 0.18]{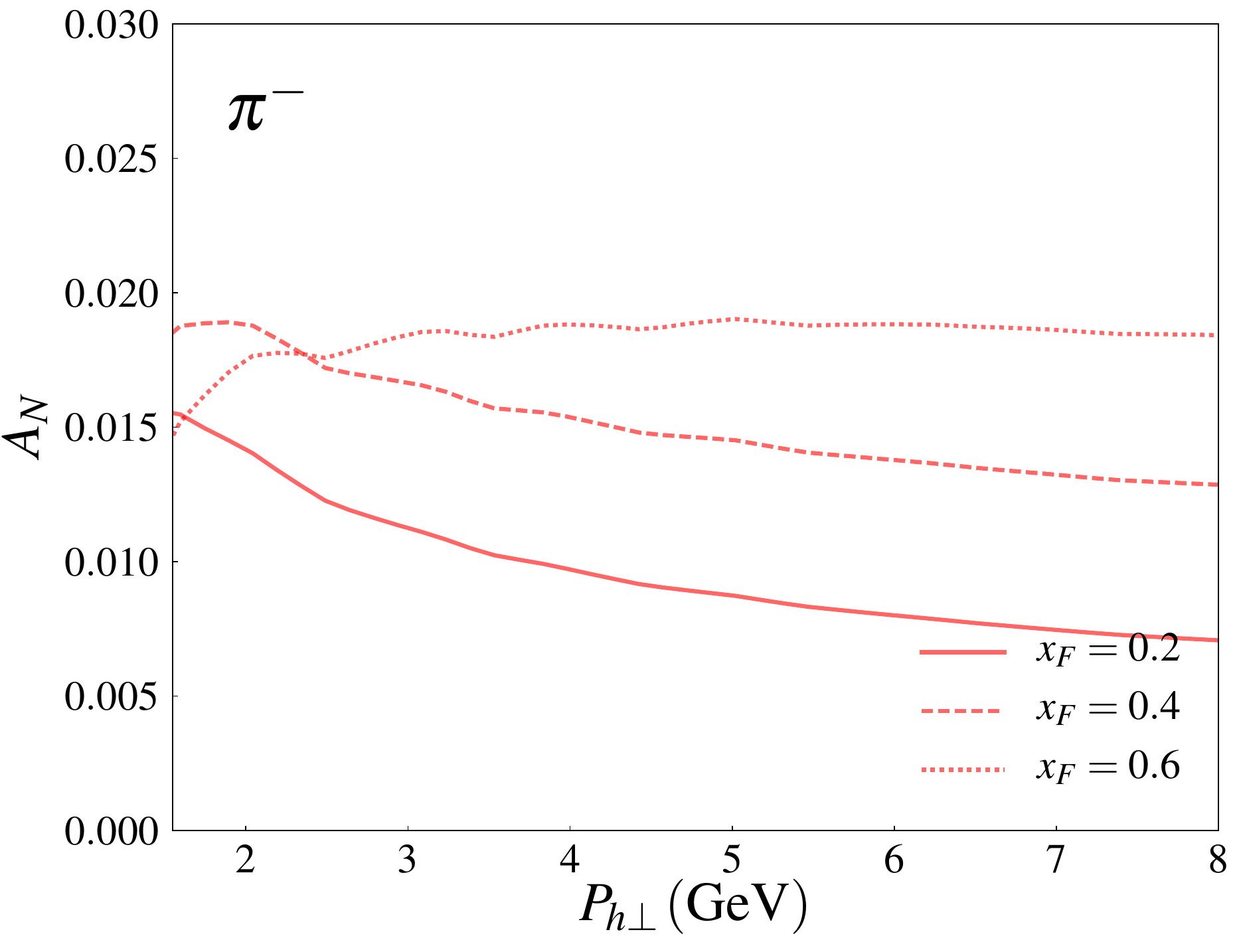}
      \includegraphics[scale = 0.18]{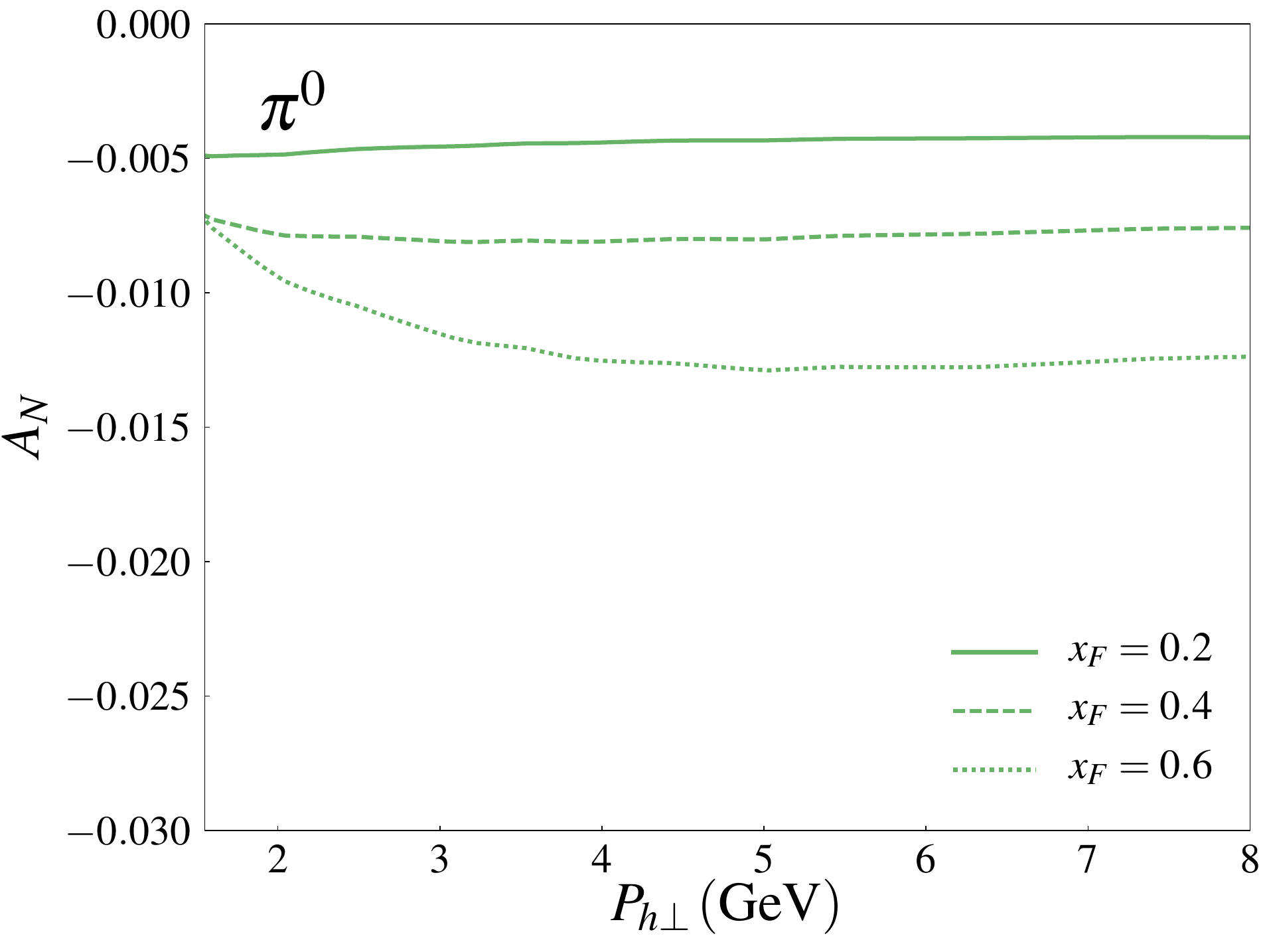}
  \end{center}
  \caption{
  $A_N$ as a function of $P_{h\perp}$ for $\pi^+$,  $\pi^-$ and $\pi^0$ and for several values of $x_F$ at $\sqrt{s} = 200$ GeV.}
  \label{fig:pifor}
\end{figure}

In this Section we show the numerical results of our calculation.
We consider the following quantity
\begin{equation}
A_N \equiv \frac{1}{2}\frac{d\Delta \sigma_{p^\uparrow A} - d\Delta\sigma_{p^\downarrow A}}{d\sigma_{pA}}
\label{eq:AN}
\end{equation}
where  $d\sigma_{pA}$ is given by \eqref{eq:xqqgunpol} (with appropriate convolutions with the photon flux), and $d\Delta\sigma_{p^\downarrow A} = -d\Delta\sigma_{p^\uparrow A}$. We also choose the standard convention $\sin(\Phi_s - \chi) = +1$, so that with an incoming proton in the $+z$ direction and spin in the $y$ direction $d\Delta\sigma_{p^\uparrow A}$ ($d\Delta\sigma_{p^\downarrow A}=-d\Delta\sigma_{p^\uparrow A}$) is the cross section for hadron emission in the $+x$ ($-x$), or left (right), direction. 
The result will be plotted as a function of the Feynman-$x_F$ variable
\begin{equation}
x_F \equiv \frac{2 P^3_h}{\sqrt{s}} = \frac{2 P_{h\perp} \sinh y_h}{\sqrt{s}}\,.
\end{equation}

A recent analysis of the experimental data for $p^\uparrow p \to \pi X$ in the forward region ($x_F > 0$) \cite{Kanazawa:2014dca,Gamberg:2017gle} indicates that the dominant contribution to single spin asymmetry originates from the twist-3 fragmentation contribution, while the quark-gluon contribution and the tri-gluon contribution are small. We thus only include the  fragmentation contribution \eqref{eq:xsf2} in the numerator of (\ref{eq:AN}) and 
 use the recent extraction of the twist-3 fragmentation functions (and transversity) from \cite{Kang:2015msa}. 
Our calculation scheme follows \cite{Kang:2015msa}.
For the unpolarized PDFs we use the central CTEQ10 set \cite{Lai:2010vv}. We use the central DSSV unpolarized fragmentation functions $D_{\pi/a}(z,\mu^2)$ \cite{deFlorian:2014xna} at the scale $\mu^2$.
We numerically solve the QCD evolution equations for the twist-3 fragmentation functions and the transversity. While the evolution equation for $h_1^a(x,\mu^2)$ is rather simple, evolution of $H_1^{\perp (1),a}(z,\mu^2)$ includes also a contribution from $\mathrm{Im} \hat{H}^a_{FU}(z,z',\mu^2)$. Following the suggestion in \cite{Kang:2015msa} we use a simplified setup assuming the contribution from $H_1^{\perp (1),a}(z,\mu^2)$ in the evolution equation is small. In this approximation, the evolution equation for $z H_1^{\perp (1),a}(z,\mu^2)$ becomes identical to the one for $h_1^a(x,\mu^2)$. Using the initial conditions\footnote{The initial condition in \cite{Kang:2015msa} for transversity is parametrized in terms of quark helicity distribution and we utilize the NLO DSSV extraction \cite{deFlorian:2009vb} as in \cite{Kang:2015msa}.} from \cite{Kang:2015msa} we have solved the resulting evolution equations using the numerical method from \cite{Hirai:1997mm}. The obtained results agree with \cite{Kang:2015msa}\footnote{Ref.~\cite{Kang:2015msa} presented numerical solutions for the evolution of the twist-3 fragmentation function $\hat{H}^{(3),a}(z,\mu^2) = -2z M_h H_1^{\perp (1),a}(z,\mu^2)$, where we used Eq.~\eqref{eq:Hs} in the last equality.}.
We also use the Wilczek-Wandzura approximation \cite{Wandzura:1977qf}: $H^a(z,\mu^2) = -2z H_1^{\perp (1),a}(z,\mu^2)$, as in Ref.~\cite{Gamberg:2017gle} so that the final numerical result for the cross section is completely determined by $H_1^{\perp (1),a}(z,\mu^2)$. Finally, we have set $\mu^2 = P_{h\perp}^2$ and used $M_N = 1$ GeV, $M_h = 0.14$ GeV and $R_p = 1$ fm, $R_A = 6$ fm appropriate for a $p^\uparrow Au$ collision.

\begin{figure}
  \begin{center}
  \includegraphics[scale = 0.25]{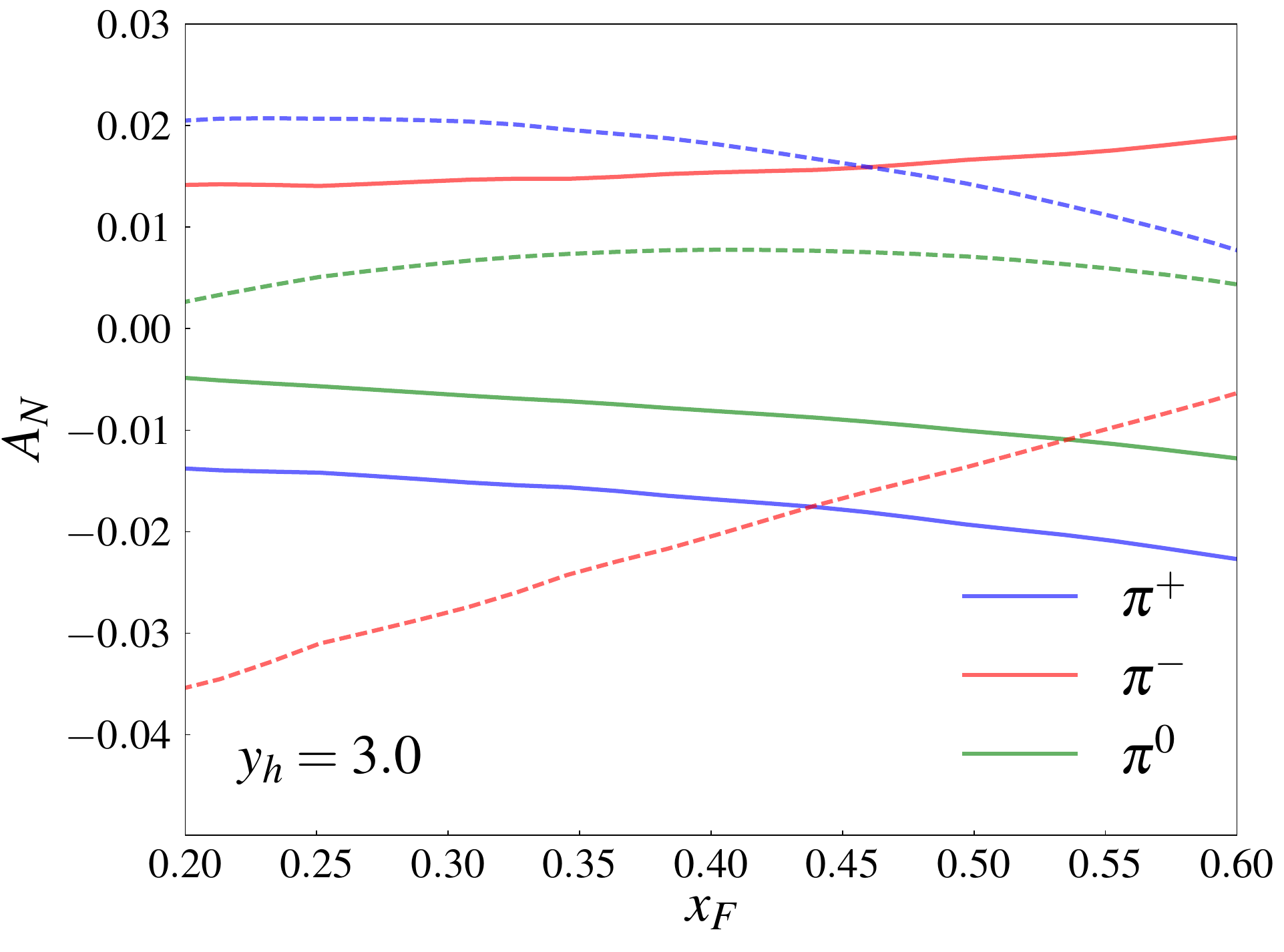}
  \end{center}
  \caption{
  $A_N$ as a function of $x_F$ for $\pi^+$, $\pi^-$ and $\pi^0$ at $y_h = 3.0$ and $\sqrt{s} = 200$ GeV. 
  The total contribution and the contribution from the third term in Eq.~\eqref{eq:xsf3} are shown by full and dashed lines, respectively.}
  \label{fig:pifor2}
\end{figure}

The numerical evaluations for $A_N$, as defined in \eqref{eq:AN}, are shown on Fig.~\ref{fig:pifor} as a function of $P_{h\perp}$, and for fixed $x_F$, while on Fig.~\ref{fig:pifor2}, in full lines, we show the results as a function of $x_F$ for fixed $y_h$, for $\pi^{\pm}$ and $\pi^0$ at $\sqrt{s} = 200$ GeV. We have used the fragmentation contribution to calculate the polarized cross section (as described in the previous paragraph), while the unpolarized cross section is given by the first two contributions in Eq.~\eqref{eq:xqqgunpol}.

The results on Fig.~\ref{fig:pifor} demonstrate $A_N$ of the order of a few percent with $A_N$ negative for $\pi^+$ and $\pi^0$ and positive for $\pi^-$. The largest $A_N$ is found for $\pi^+$, while for $\pi^0$ it is about a factor of two smaller than for $\pi^+$. Additionally, $A_N$ shows a mild $P_{h\perp}$ dependence and an overall increase in magnitude with $x_F$ for all charges as seen from Fig.~\ref{fig:pifor} and also summarized on Fig.~\ref{fig:pifor2} (full lines).
The results bear some qualitative similarity to the typical SIDIS results, see e.~g. Fig.~5 in \cite{Parsamyan:2015nnl}. The monotonic increase of $A_N$ with $x_F$ is typical also for $p^\uparrow p$ collisions, however the magnitude of $A_N$ in $p^\uparrow p$ is about 10 percent.
On the other hand, in $p^\uparrow p$ collisions $A_N >0$  for $\pi^+$ and $\pi^0$, while $A_N <0$ for $\pi^-$, which is the opposite to the case of UPC. This difference is due to the additional $t$-channel gluon exchanges in the $qg$ contribution, dominating the high energy $p p$ cross section, while such analogous contributions are absent in the $q\gamma$ contribution to the UPC cross section.

To expand on this sign difference, we use the QCD equation of motion relation (first Eq.~\eqref{eq:ft3}) to write the fragmentation contribution to the polarized cross section Eq.~\eqref{eq:xsf}  as
\begin{equation}
\begin{split}
& \frac{d\Delta\sigma_{\rm frag}}{d^2 P_{h\perp} d y_h} = \frac{8 M_h P_{h\perp}\alpha_{em}\alpha_s}{S}\sin(\Phi_S - \chi)\int_{x_{\rm min}}^1 \frac{dx}{x}\int_{z_{\rm min}}^1 \frac{dz}{z^3}\delta\left(\hats + \hatt + \hatu\right)\sum_{a} e_a^2 h_1^a(x,\mu^2)\\
& \times\Bigg[\left(H_1^{\perp (1),a}(z,\mu^2) - z \frac{d H_1^{\perp (1),a}(z,\mu^2)}{dz}\right)\Delta\hat{\sigma}_{H_1^\perp} + \frac{H^a(z,\mu^2)}{z}\Delta \hat{\sigma}_H\\
& + \frac{2}{z}\int_z^\infty \frac{dz'}{z'^2}P\left(\frac{1}{(1/z - 1/z')^2}\right)\mathrm{Im} \hat{H}^a_{FU}(z,z',\mu^2)\Delta \hat{\sigma}_{\hat{H}_{FU}}\Bigg] \,,
\end{split}
\label{eq:xsf3}
\end{equation}
where
\begin{equation}
\Delta \hat{\sigma}_{H_1^\perp} = 2\Delta \hat{\sigma}_H \equiv 8C_F \frac{1}{\hatu}\,,\qquad \Delta \hat{\sigma}_{\hat{H}_{FU}} \equiv \frac{4}{\hatu}\left(\frac{1}{N_c}\frac{\hats}{\hatt} + N_c\right)\,.
\end{equation}
Eq.~\eqref{eq:xsf3} has  an identical structure (apart from the overall numerical coefficient and the explicit form of the hard factors) as in $p^\uparrow p$ collisions, see Eq.~(9) in \cite{Gamberg:2017gle}. Ref.~\cite{Gamberg:2017gle} found that the third term in their Eq.~(9), gives the largest contribution to $A_N$. On Fig.~\ref{fig:pifor2} we plot the corresponding contribution in $p^\uparrow A$ UPC, which is the third term in Eq.~\eqref{eq:xsf3}. Unlike in the $p^\uparrow p$ case, it is seen from Fig.~\ref{fig:pifor2} that this term contributes to $A_N$ with an opposite sign from the remaining terms in the cross section and also that it gets completely overcompensated by the remaining terms.

\begin{figure}
  \begin{center}
  \includegraphics[scale = 0.25]{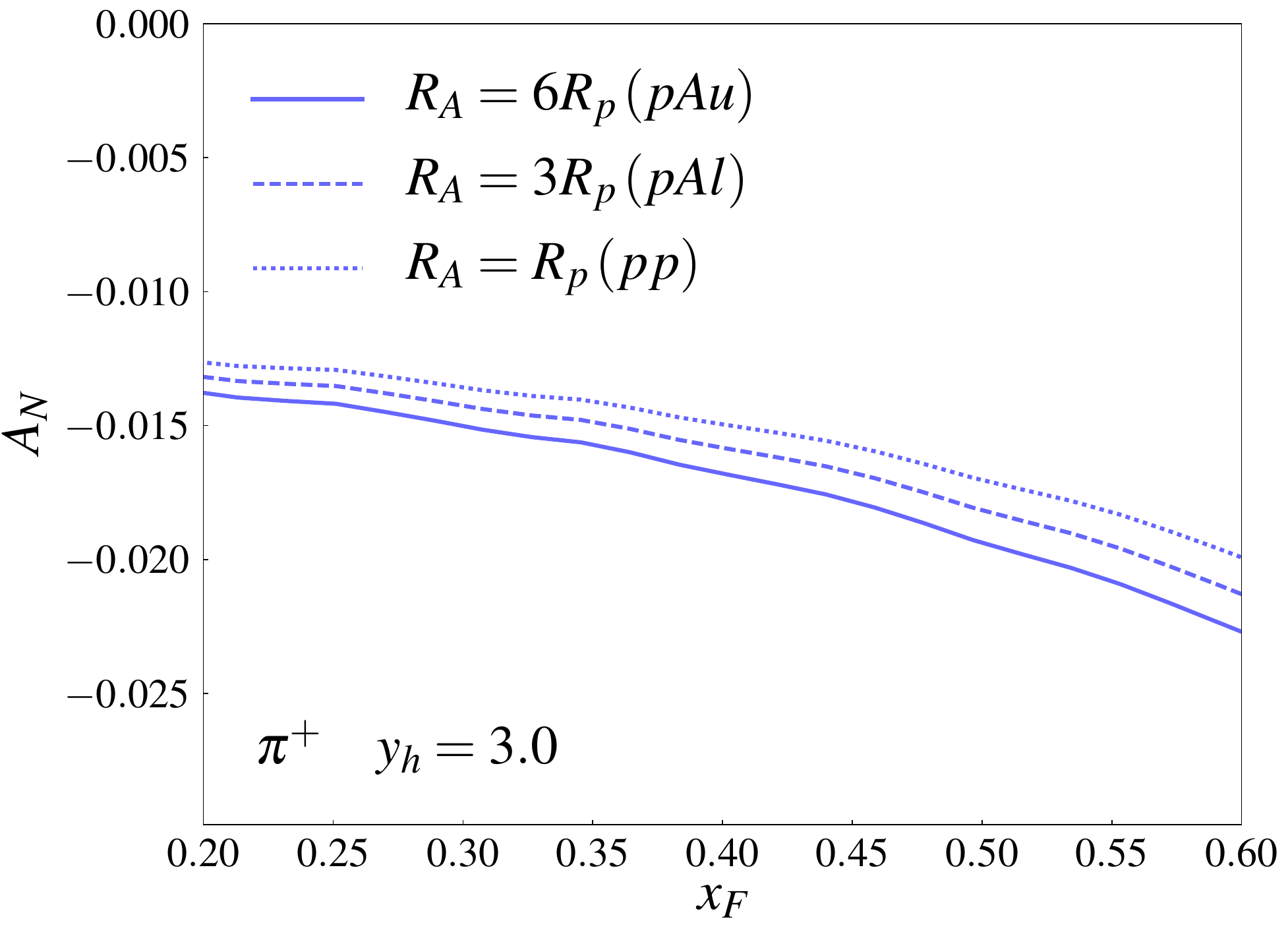}
  \end{center}
  \caption{$A_N$ as a function of $x_F$ for $\pi^+$ at $y_h = 3.0$ and $\sqrt{s} = 200$ GeV. The three different curves stand for the choices for the radii of the nuclei $R_A$.}
  \label{fig:nuc}
\end{figure}

Lastly, we study the nuclear dependence of $A_N$. On Fig.~\ref{fig:nuc} we focus on $\pi^+$ and reduce $R_A$ from the original value $R_A = 6 R_p$ to $R_A = 3 R_p$ and $R_A = R_p$, with the last two choices appropriate for a $p^\uparrow Al$ and $p^\uparrow p$ UPCs, respectively. We find that $A_N$ decreases in magnitude by decreasing $R_A$. The quantitative effect is, however, rather small, as the nuclear dependence, to a large extent, cancels in the ratio.

\section{Conclusions}
\label{sec:conc}

We have studied single spin asymmetries in ultra-peripheral $p^\uparrow A$ collisions. We have provided a set of formulas for unpolarized and transversely polarized $p^\uparrow A \to \pi X$ cross section. The final expressions for the polarized cross section, contained in Eqs.~\eqref{eq:dsigV}, \eqref{eq:dsigA}, \eqref{eq:3g} and \eqref{eq:xsf2}, describe the twist-3 quark-gluon, twist-3 gluon and the twist-3 fragmentation contribution, respectively, thus fully accounting for all the known sources of single spin asymmetries within the twist-3 collinear framework. While these results are straightforward adaptations of the $Q^2 \to 0$ limit in SIDIS, the explicit expressions turn out to be rather simple, demonstrating the advantage of the UPC channel as a new probe of the distribution functions of the polarized proton.

In order to illustrate the potential of the UPC channel we have performed a numerical calculation of the SSA in the forward region at $\sqrt{s} = 200$ GeV. We assumed that the dominant source to the asymmetry is the twist-3 fragmentation contribution. While this assumption seems to be supported by the calculation in $p^\uparrow p$ collisions \cite{Kanazawa:2014dca,Gamberg:2017gle}, we can check it with the UPC channel.
Using the central values of the most recent extractions of the transversity distribution and the twist-3 fragmentation functions from \cite{Kang:2015msa} our findings point to an asymmetry of the order of a few percent. A notable feature is that $A_N$ for $\pi^+$ ($\pi^-$) is negative (positive), in contrast to the $pp$ case where it is positive (negative) and large. As RHIC is presently conducting polarized $p A$ collisions, this prediction can be tested by tagging the UPC events. It would be interesting to perform a more complete calculation by adding also the twist-3 quark-gluon contribution. Nevertheless, we hope these first results are encouraging enough for future experimental extractions of SSA in UPC.

\acknowledgments{We are grateful to Daniel Pitonyak and Alexei Prokudin for help. S.~B. is supported by a JSPS postdoctoral fellowship for foreign researchers under Grant No.~17F17323. S.~B. also acknowledges HRZZ Grant No. 8799 for computational resources.}

\appendix

\section{Twist-3 quark-gluon and twist-3 gluon contributions}
\label{app:qg3g}

In this appendix we collect the main formulas for the twist-3 quark-gluon and the twist-3 gluon contribution to the cross section of the $p^\uparrow\gamma$ collision.
The relevant formula for the twist-3 quark-gluon contribution to the SIDIS cross section was calculated in \cite{Eguchi:2006qz,Eguchi:2006mc}, and below we follow the notation from \cite{Eguchi:2006mc}. We first concentrate on the contribution from the Qiu-Sterman function $G_F(x,y,\mu^2)$, which is also the first $k_\perp$-moment of the Sivers function, see Sec.~2.1 in \cite{Eguchi:2006mc} for the explicit definition.
The relevant cross section, see Eq.~(80) (Eq.~(85)) in \cite{Eguchi:2006mc} for the quark (gluon) fragmentation channel, in the $Q^2 \to 0$ limit is
\begin{equation}
\begin{split}
\frac{d\Delta\sigma_{q\qbar g}^V}{{d^2 P_{h\perp} dy_h}} &= -\frac{2\pi M_N P_{h\perp} \alpha_{em}\alpha_s}{S}\sin(\Phi_S - \chi)\int_{x_{\rm min}}^1 \frac{dx}{x}\int_{z_{\rm min}}^1 \frac{dz}{z^3}\delta\left(\hats+\hatt+\hatu\right)\\
&\times\sum_{a}e_a^2 \Bigg\{\left[\delta_a D_{h/q}(z,\mu^2)\hat{\sigma}^{Vq}_{D1} + D_{h/g}(z,\mu^2)\hat{\sigma}^{Vg}_{D1}\right] \left[x\frac{d}{dx}G^a_F(x,x,\mu^2) - G^a_F(x,x,\mu^2)\right]\\
& + \delta_a \left[D_{h/q}(z,\mu^2)\left(\hat{\sigma}^{Vq}_{H1} + \hat{\sigma}^{Vq}_{F1}\right) + D_{h/g}(z,\mu^2)\left(\hat{\sigma}^{Vg}_{H1} + \hat{\sigma}^{Vg}_{F1}\right)\right]G^a_F(0,x,\mu^2)\Bigg\}\,,
\label{eq:dsigV}
\end{split}
\end{equation}
where $\delta_a = 1$ for quark and $\delta_a = -1$ for antiquark.
The relevant hard factors, see Eqs.~(81)-(84) and Eqs.~(86)-(89) in \cite{Eguchi:2006mc} (for convenience, we have divided the original expressions by $q_T \hatz$, while keeping the same notation), in the $Q^2 \to 0$ limit are
\begin{equation}
\begin{split}
&\hat{\sigma}^{Vq}_{D1} = -\hat{\sigma}^{Vq}_{G1} =  \frac{4}{N_c}\frac{1}{\hats}\frac{1 + (1-\hatz)^2}{\hatz(1-\hatz)^2} = -\frac{4}{N_c}\frac{\hats}{\hatt\hatu}\left(\frac{\hats}{\hatu} + \frac{\hatu}{\hats}\right)\,,\\
&\hat{\sigma}_{H1}^{Vq} = \frac{8}{\hats}\frac{1}{\hatz(1-\hatz)^2}\left( C_F \hatz + \frac{1}{2N_c}\right) = \frac{4\hats}{\hatt\hatu}\left(N_c \frac{\hatt}{\hatu}+\frac{1}{N_c}\right)\,,\\
&\hat{\sigma}_{F1}^{Vq} = \frac{4}{N_c}\frac{1}{\hatt}\,,
\end{split}
\end{equation}
\begin{equation}
\begin{split}
&\hat{\sigma}^{Vg}_{D1} = - \hat{\sigma}^{Vg}_{G1} = -\frac{4N_c}{\hats}\frac{1+\hatz^2}{\hatz^2(1-\hatz)} = -4N_c\frac{\hats}{\hatt\hatu}\left(\frac{\hats}{\hatt} + \frac{\hatt}{\hats}\right)\,,\\
&\hat{\sigma}^{Vg}_{H1} = \frac{8}{\hats}\frac{1}{\hatz^2(1-\hatz)}\left[C_F(\hatz - 1)-\frac{1}{2N_c}\right] = -\frac{4}{N_c}\frac{\hats}{\hatt\hatu}\left(N_c^2 \frac{\hatu}{\hatt} + 1\right)\,,\\
&\hat{\sigma}^{Vg}_{F1} = \frac{4}{N_c}\frac{1}{\hats}\frac{1}{(1-\hatz)} = -\frac{4}{N_c}\frac{1}{\hatu}\,.\\
\end{split}
\end{equation}

Next, we write the contribution from the Qiu-Sterman function $\tilde{G}_F(x,y)$. The relevant cross section, see Eq.~(90) in \cite{Eguchi:2006mc}, in the $Q^2 \to 0$ limit is
\begin{equation}
\begin{split}
\frac{d\Delta\sigma_{q\qbar g}^{A}}{d^2 P_{h\perp} dy_h} &= -\frac{2\pi M_N P_{h\perp}\alpha_{em}\alpha_s}{S}\sin(\Phi_S - \chi)\int_{x_{\rm min}}^1 \frac{dx}{x}\int_{z_{\rm min}}^1 \frac{dz}{z^3}\delta\left(\hats+\hatt+\hatu\right)\\
&\times\sum_{a}e_a^2\delta_a\left[\left(\hat{\sigma}^{Aq}_{H1} + \hat{\sigma}^{Aq}_{F1}\right) D_{h/a}(z,\mu^2) + \left(\hat{\sigma}^{Ag}_{H1} + \hat{\sigma}^{Ag}_{F1}\right)D_{h/g}(z,\mu^2)\right]\tilde{G}^a_F(0,x,\mu^2)\,.
\end{split}
\label{eq:dsigA}
\end{equation}
where the relevant hard factors, see Eqs.~(91)-(94) in \cite{Eguchi:2006mc} (divided by $q_T\hatz$ in our notation), in the $Q^2 \to 0$ limit are
\begin{equation}
\begin{split}
&\hat{\sigma}^{Aq}_{H1} = \hat{\sigma}^{Vq}_{H1}\,,\\
&\hat{\sigma}^{Aq}_{F1} = \frac{\hatz^2}{(1-\hatz)^2}\hat{\sigma}^{Vq}_{F1} = \frac{\hatt^2}{\hatu^2}\hat{\sigma}^{Vq}_{F1}\,,\\
& \hat{\sigma}^{Ag}_{H1} = \hat{\sigma}^{Vg}_{H1}\,,\\
&\hat{\sigma}^{Ag}_{F1} = \hat{\sigma}^{Vg}_{F1}\,.
\end{split}
\end{equation}
The complete contribution is obtained by summing \eqref{eq:dsigV} and \eqref{eq:dsigA}.

The twist-3 gluon contribution to the SIDIS cross section was calculated in \cite{Kang:2008qh,Beppu:2010qn} and we use the notation from \cite{Beppu:2010qn}. The formula takes into account the twist-3 distributions $N(x,y,\mu^2)$ and $O(x,y,\mu^2)$.
Taking the $Q^2 \to 0$ and $m_c^2 \to 0$ limit in Eq.~(68) in \cite{Beppu:2010qn}, we find
\begin{equation}
\begin{split}
& \frac{d\Delta\sigma_{ggg}}{d^2 P_{h\perp}dy_h} =  -\frac{2\pi M_N P_{h\perp}\alpha_{em}\alpha_s}{S}\sin(\Phi_S-\chi)\int_{x_{\rm min}}^1 \frac{dx}{x}\int_{z_{\rm min}}^1 \frac{dz}{z^3}\delta\left(\hats + \hatt + \hatu\right)\sum_{a} e_a^2 D_{h/a}(z,\mu^2)\\
&\times\left\{\delta_a\left[\frac{d}{dx}O(x,\mu^2) - \frac{2 O(x,\mu^2)}{x} \right] + \left[\frac{d}{dx}N(x,\mu^2) - \frac{2 N(x,\mu^2)}{x} \right]\right\}\Delta \hat{\sigma}_g\,,
\label{eq:3g}
\end{split}
\end{equation}
where $\delta_a = +1(-1)$ for quark (antiquark) and where the relevant hard factors, see Eqs.~(71)-(74) in \cite{Beppu:2010qn} (divided by $q_T\hatz$ in our notation), in the $Q^2 \to 0$ limit are
\begin{equation}
\Delta \hat{\sigma}_g = \frac{8}{\hats}\frac{(1-\hatz)^2 + \hatz^2}{\hatz^2(1-\hatz)^2} = \frac{8\hats}{\hatt\hatu}\left(\frac{\hatu}{\hatt} + \frac{\hatt}{\hatu}\right)\,.
\end{equation}
We also introduced the notation $O(x,\mu^2) \equiv O(x,x,\mu^2) + O(x,0,\mu^2)$, $N(x,\mu^2) \equiv N(x,x,\mu^2) - N(x,0,\mu^2)$ 
where  $O(x,y,\mu^2)$ ($N(x,y,\mu^2)$) is the $C$-odd ($C$-even) twist-3 gluon distribution. It is interesting to note that in the $Q^2\to 0$ limit the cross section depends only on these specific linear combinations. This is not the case in  generic situations \cite{Beppu:2010qn}.   While $N(x,y,\mu^2)$ is related to the first $k_\perp$-moment of the gluon Sivers function, $O(x,y,\mu^2)$ is related to the polarized odderon \cite{Zhou:2013gsa}.
Eq.~\eqref{eq:3g} could be used to directly extract $O(x,\mu^2)$ in UPC with backward jet production.

\bibliographystyle{h-physrev}
\bibliography{references}

\end{document}